\UseRawInputEncoding
\documentclass[preprintnumbers, floatfix, twocolumn,
preprintnumbers, PRD, superscriptaddress,nofootinbib]{revtex4}
\usepackage{eurosym}
\usepackage{amsfonts}
\usepackage{amsmath}
\usepackage{amssymb,epsf}
\usepackage{latexsym}
\usepackage{graphicx,epsfig}
\usepackage{amssymb}
\usepackage{subfigure}
\usepackage{epstopdf}
\usepackage[colorlinks=true,citecolor=blue,linkcolor=blue,urlcolor=black]{hyperref}
\usepackage{color}
\usepackage{float}

\graphicspath{{Images/}}

\renewcommand{\&}{\textup{\symbol{`\&}}}

\begin{document}

\title{Growth of Perturbations in Tsallis and Barrow Cosmology} 

\author{Ahmad Sheykhi}
\email{asheykhi@shirazu.ac.ir} \affiliation{Department of Physics,
College of Sciences, Shiraz University, Shiraz 71454, Iran}
\affiliation{Biruni Observatory, College of Sciences, Shiraz
University, Shiraz 71454, Iran}

\author{Bita Farsi}
\email{Bita.Farsi@shirazu.ac.ir} \affiliation{Department of
Physics, College of Sciences, Shiraz University, Shiraz 71454,
Iran}

\begin{abstract}
We disclose the effects of the entropic corrections to the
Friedmann equations on the growth of perturbations in the early
stages of the universe. We consider two types of corrections to
the area law of entropy, known as Tsallis and Barrow entropies.
Using these corrections to entropy, we derive the modified
Friedmann equations and explore the growth of perturbations in a
flat universe filled with dark matter (DM) and cosmological
constant. We employ the spherically symmetric collapse formalism
and work in the liner regime for the perturbations. Interestingly
enough, we find out that the profile of density contrast quite
differs from the standard cosmology in Tsallis and barrow
cosmology. We observe that the growth rate of matter perturbations
crucially depend on the values of Tsallis and Barrow parameters.
With increasing these correction parameters to the entropy, the
total density contrast increases as well. This implies that
perturbations grow up faster in a universe with modified entropy
corrected Friedmann equations.
\end{abstract}
\maketitle
\section{Introduction \label{Intro}}
One of the main challenges of the modern cosmology is
understanding the origin and physics of the growth of
perturbations in the early stages of the universe. Indeed, these
perturbations eventually lead to the large scale structures such
as galaxies and clusters of galaxy. It is a general belief that
the large scalae structures of the universe originate from
gravitational instability that amplifies very small initial
density fluctuations during the universe evolution. Such
fluctuations then grow slowly over time until they get robust
enough to be detached from the background expansion. Finally, they
collapse into gravitationally-bound systems such as galaxies and
clusters of galaxy. In fact, the primordial collapsed regions
serve as the initial cosmic seeds for which the large scale
structures are developed \cite{Peebles,White}. This gives enough
motivations to study the perturbations of matter and DE at the
early stages of the universe in the linear and non-linear regimes.
It was argued that the perturbations of DE may form halo
structures which can influence the DM collapsed region nonlinearly
\cite{Abramo}. Studies of the mutual interactions between dark
sectors of the universe, in the perturbation level, can shed light
to understand the nature of both dark components of the universe.
Influence of the DE on structure formation both at the background
level (no fluctuations) and the perturbed level has been widely
explored in the literatures
\cite{Abramo2,Nunes,Liberato,Dutta,Amendola}. An appropriate
approach to investigate the growth of perturbations and structure
formation is the so called Top-Hat Spherical Collapse (SC) model
\cite{Abramo}. In this approach one considers a uniform and
spherical symmetric perturbation in an expanding background and
describes the growth of perturbations in a spherical region using
the same Friedmann equations for the underlying theory of gravity
\cite{Planelles,Ziaei1,Ziaei2,Farsi}.

According to thermodynamics-gravity conjecture
\cite{Jacobson,Padmanabhan1,Padmanabhan2,Paranjape,Kothawala,Eling,Padmanabhan3,Frolov,Calcagni},
one can write the Friedmann equations, in any gravity theory, in
the from of the first law of thermodynamics, $dE = T_{h}dS_{h} +W
dV$, on the apparent horizon and vice versa
\cite{Cai1,Cai2,Akbar1,Akbar2,Akbar3,
Sheykhi1,Sheykhi2,Sheykhi3,Sheykhi4,Sheykhi5,SheyECFE,Sheyem}. For
this purpose, one should pick up the entropy expression of the
black hole in each gravity theory and replace the black hole
horizon radius $r{+}$ by the apparent horizon radius
$\tilde{r}_{A}$. While it is more convenient to apply the
Bekenstein-Hawking area law relation defined as $S_{h}=A/(4G)$ for
the black hole entropy (with the black hole horizon area $A=4 \pi
r_{+}^{2}$), it should be noted that, there are several types of
corrections to the area law of entropy. Two possible corrections
occur from the inclusion of quantum effects are known as
logarithmic and power-law corrections. Logarithmic corrections,
arises from the loop quantum gravity due to thermal equilibrium
fluctuations and quantum fluctuations
\cite{Das,Ashtekar,Zhang,Banerjee,SheyLog} and the power-law
corrections appears in dealing with the entanglement of quantum
fields inside and outside the horizon
\cite{Das2,Radicella,SheyPL}.

Another type of correction to the area law of entropy comes from
the fact that, in divergent partition function systems like
gravitational systems, Boltzmann-Gibbs additive entropy should be
generalized to non-additive entropy
\cite{Wilk,Gibbs,RNunes,Tsallis1}. Based on this, and using the
statistical arguments, Tsallis and Cirto \cite{Tsallis2} argued
that the microscopic mathematical expression of the
thermodynamical entropy of a black hole does not obey the area law
and get modified as $S_{h} \sim A^{\beta}$, where $\beta$ known as
nonextensive parameter, which quantifies the degree of
nonextensivity of the system \cite{Tsallis2}. Tsallis definition
of entropy plays a crucial role in studying the gravitational and
cosmological systems in the framework of the generalized
statistical mechanics \cite{Czinner,Sayahian,Moradpour}. Some
phenomenological aspects as well as some observational constraints
on the modified Friedmann equations based on Tsallis entropy were
explored in \cite{Asghari}. It was argued that this model is
compatible with observations \cite{Asghari}. Modified cosmology
through Tsallis holographic DE (THDE) have been also explored
\cite{Tavayef,Abd,Bram,Huang,Bhattacharjee}. It was shown that
depending on the value of $\beta$ parameter, THDE model can
explain the current accelerated cosmic expansion and alleviate the
age problem \cite{Huang}.

Recently, Barrow suggested a fractal structure for the black hole
horizon and argued that the area of the horizon may increase due
to the quantum-gravitational deformation \cite{Barrow}. As a
result, the area law of entropy get modified as $S_{h} \sim A^{
1+\Delta/2} $ where $\Delta$ quantifies the quantum-gravitational
deformation. The modified cosmological field equations based on
Barrow entropy have been explored in \cite{Sheykhi2,Saridakis1}.
Later on, new developments have appeared in the literature aiming
to test the performance of the Barrow entropy in the cosmological
framework \cite{Saridakis2}. The validity and the constraints
imposed by the generalized second law of thermodynamics, including
the matter-energy content and the horizon entropy, have been
investigated in \cite{Saridakis2}. Additionally, the Barrow
holographic DE (BHDE) model was proposed in
\cite{Saridakis3,Sri,Adh,Oliv} and has been tested against the
latest cosmological data in \cite{Anagnostopoulos,Dabrowski}. It
was found that BHDE describes very efficiently the late
accelerated expansion of the universe, having additionally the
correct asymptotic behavior \cite{Mamon}. It is worth noting that
although the Barrow entropy resembles Tsallis entropy, however,
the physical motivation and origin of them are completely
different. Indeed, Tsallis non-additive entropy correction is
motivated by generalizing standard thermodynamics to a
non-extensive one, while Barrow correction to entropy originates
from intricate, fractal structure on horizon induced by
quantum-gravitational effects.

Our aim in this paper is to disclose the influences of the Tsallis
and Barrow entropy corrections on the growth of perturbations in
the early stage of the universe. For this purpose, using
thermodynamics-gravity conjecture we extract the modified
Friedmann equations based on Tsallis/Barrow entropy
\cite{Sheykhi1,Sheykhi2}. We then employ the SC approach, and
investigate the evolution of DM and DE perturbations in the
framework of the Tsallis and Barrow cosmology.

The paper is structured as follows. In section \ref{Fri}, we
provide a review on the modified Friedmann equations in Tsallis
and Barrow cosmology. In section \ref{GTE}, we examine the growth
of matter perturbations in Tsallis cosmology with using the
SC approach. In section \ref{GBE}, we explore the growth of
perturbations in the background of Barrow cosmology. We finish our
paper with concluding remarks in section \ref{Con}.
\section{Modified Friedmann equations in Tsallis and Barrow Cosmology\label{Fri}}
We start with a spatially homogeneous and isotropic spacetime with
line elements
\begin{equation}
ds^2={h}_{\mu \nu}dx^{\mu}
dx^{\nu}+\tilde{r}^2(d\theta^2+\sin^2\theta d\phi^2),
\end{equation}
where $\tilde{r}=a(t)r$, $x^0=t, x^1=r$, and $h_{\mu \nu}$=diag
$(-1, a^2/(1-kr^2))$ represents the two dimensional metric. The
open, flat, and closed universes corresponds to $k = -1,0, 1$,
respectively. We assume our universe is bounded by an apparent
horizon, which its radius is \cite{Sheykhi3}
\begin{equation}
\label{radius} \tilde{r}_A=\frac{1}{\sqrt{H^2+k/a^2}}.
\end{equation}
The apparent horizon is a suitable boundary for which the first
and second law of thermodynamics hold on it. The temperature
associated with the apparent horizon has the form
\cite{Cai2,Sheykhi3}
\begin{equation}\label{T}
T_h=-\frac{1}{2 \pi \tilde r_A}\left(1-\frac{\dot {\tilde
r}_A}{2H\tilde r_A}\right).
\end{equation}
We further assume the energy content of the universe is in the
form of perfect fluid,
\begin{equation}\label{EMT}
T_{\mu\nu}=(\rho+p)u_{\mu}u_{\nu}+pg_{\mu\nu},
\end{equation}
where $\rho$ and $p$ are, respectively, the energy density and
pressure of the matter inside the Universe. The total energy
content of the Universe satisfies the conservation equation,
$\nabla_{\mu}T^{\mu\nu}=0$, which yields the continuity equation,
\begin{equation}\label{Cont}
\dot{\rho}+3H(\rho+p)=0.
\end{equation}
In the above relation $H=\dot{a}/a$ stands for the Hubble
parameter which measures the rate of the universe expansion. In an
expanding universe, as a thermodynamic system, we need to define a
work term in the first law of thermodynamics. The work density for
an expanding universe may be obtained as \cite{Hay}
\begin{equation}\label{Work}
W=-\frac{1}{2} T^{\mu\nu}h_{\mu\nu},\ \ \Rightarrow \ \
W=\frac{1}{2}(\rho-p).
\end{equation}
We further assume the first law of thermodynamics on the apparent
horizon is satisfied and has the form
\begin{equation}\label{FL}
dE = T_h dS_h + WdV,
\end{equation}
where $V=\frac{4\pi}{3}\tilde{r}_{A}^{3}$ is the volume enveloped
by a 3-dimensional sphere, and $T_{h}$ and $W$ are given by
(\ref{T}) and (\ref{Work}). In the above expression, $S_{h}$
stands for the entropy associated with the apparent horizon. The
total energy of the universe is $E=\rho V$, which after taking
differential and using the continuity equation (\ref{Cont}), leads
to
\begin{equation}
\label{dE} dE=4\pi\tilde {r}_{A}^{2}\rho d\tilde {r}_{A}-4\pi H
\tilde{r}_{A}^{3}(\rho+p) dt.
\end{equation}
In order to derive the Friedmann equations from the first law of
thermodynamics (\ref{FL}), we need to define the entropy
expression associated with the boundary of the Universe. In this
paper, we consider two modification for the area law, known as
Tsallis and Barrow entropy.
\subsection{Modified Friedmann equations through Tsallis entropy}
 The Tsallis entropy associated with
the boundary of the Universe is given by \cite{Tsallis2}
\begin{eqnarray}\label{ST}
S_{h}= \gamma A^{\beta},
\end{eqnarray}
where $A$ is the apparent horizon area, $\gamma$ is an unknown
constant and $\beta$ known as Tsallis parameter or nonextensive
parameter, which is a real parameter which quantifies the degree
of non-extensivity \cite{Tsallis2}. When $\beta=1$ and
$\gamma=1/(4L_p^2)$, the well-known area law of entropy is
recovered. Using Tsallis entropy in the first law of
thermodynamics on the apparent horizon, we derive the differential
form of the modified Friedmann equation \cite{Sheykhi1}
\begin{equation} \label{FriedT}
-\frac{2}{\tilde {r}_{A}^3} \left(4\pi \tilde
{r}_{A}^2\right)^{\beta-1} d\tilde {r}_{A} = \frac{2\pi }{3\gamma
\beta}d\rho.
\end{equation}
Integrating, we arrive at the modified Friedmann equation based on
Tsallis entropy \cite{Sheykhi1}
\begin{equation} \label{Fried0T}
\left(H^2+\frac{k}{a^2}\right)^{2-\beta} = \frac{8\pi L_p^2} {3}
\rho +\frac{\Lambda}{3},\end{equation} where $\Lambda$ is the
integration constant which can be interpreted as the cosmological
constant, and we have defined
\begin{equation}\label{Lp}
\gamma\equiv\frac{2-\beta }{4\beta L_p^2 } \left(4\pi
\right)^{1-\beta}.
\end{equation}
The above Friedmann equation can be rewritten as
\begin{equation} \label{Fried1T}
\left(H^2+\frac{k}{a^2}\right)^{2-\beta} = \frac{8\pi L_p^2} {3}
(\rho +\rho_{\Lambda}),
\end{equation}
where $\rho_{\Lambda}=\Lambda/(8\pi L_p^2)$ is the energy density
of the cosmological constant. Since $\gamma>0$, thus from Eq.
(\ref{Lp}) we have $\beta<2$. The second modified Friedmann
equation can be easily derived by combining the continuity
equation (\ref{Cont}) with the first Friedmann equation
(\ref{Fried1T}). We find \cite{Sheykhi1}
\begin{eqnarray}
&&(4-2\beta)
\frac{\ddot{a}}{a}\left(H^2+\frac{k}{a^2}\right)^{1-\beta}+(2\beta-1)
\left(H^2+\frac{k}{a^2}\right)^{2-\beta}\nonumber\\&&=-8\pi
L_{p}^{2} (p+p_{\Lambda}).\label{Fried2T}
\end{eqnarray}
where $p_{\Lambda}=-\Lambda/(8\pi L_p^2)$ is the vacuum
(cosmological constant) pressure. As usual, we define the density
parameters as
\begin{eqnarray}
\Omega_m=\frac{\rho_m}{\rho_{c}}, \  \  \
\Omega_{\Lambda}=\frac{\rho_{\Lambda}}{\rho_{c}}, \  \   \
\rho_{c}=\frac{3H^{4-2\beta}}{8\pi L_{p}^{2}}. \  \
\end{eqnarray}
Therefore, in terms of the density parameters, the first Friedmann
equation (\ref{Fried1T}) can be written as
\begin{eqnarray}
\Omega_m+ \Omega_{\Lambda}=(1+\Omega_k)^{2-\beta},
\end{eqnarray}
where the curvature density parameter is defined as usual,
$\Omega_k=k/(a^2H^2)$, and we have assumed $\rho=\rho_m$ by
neglecting the radiation.
\subsection{Modified Friedmann equations through Barrow entropy}
The Barrow entropy associated to the apparent horizon is given by
\cite{Barrow}
\begin{eqnarray}\label{SB}
S_{h}= \left(\frac{A}{A_{0}}\right)^{1+\Delta/2},
\end{eqnarray}
where $A$ is apparent horizon area and $A_0$ is the Planck area.
The exponent $\Delta$ ranges as $0\leq\Delta\leq1$ and stands for
the amount of the quantum-gravitational deformation effects
\cite{Barrow}. When $\Delta=0$, the area law is restored and
$A_{0}\rightarrow 4G$, while $\Delta=1$ represents the most
intricate and fractal structure of the horizon. Using the
thermodynamics-gravity conjecture, the differential form of the
Friedmann equation derived from the first law of thermodynamics,
based on Barrow entropy, is given by \cite{Sheykhi2}
\begin{equation} \label{FriedB}
-\frac{2+\Delta}{2\pi A_0
}\left(\frac{4\pi}{A_0}\right)^{\Delta/2} \frac{d\tilde
{r}_{A}}{\tilde {r}_{A}^{3-\Delta}}=
 \frac{d\rho}{3}.
\end{equation}
After integration, we find the first modified Friedmann equation
in Barrow cosmology,
\begin{equation} \label{Fried0B}
\left(H^2+\frac{k}{a^2}\right)^{1-\Delta/2} = \frac{8\pi G_{\rm
eff}}{3} \rho+\frac{\Lambda}{3},
\end{equation}
where $\Lambda$ is a constant of integration which can be
interpreted as the cosmological constant, and  $G_{\rm eff}$
stands for the effective Newtonian gravitational constant,
\begin{equation}\label{Geff}
G_{\rm eff}\equiv \frac{A_0}{4} \left(
\frac{2-\Delta}{2+\Delta}\right)\left(\frac{A_0}{4\pi
}\right)^{\Delta/2}.
\end{equation}
If we define $\rho_{\Lambda}={\Lambda}/(8\pi G_{\rm eff})$, Eq.
(\ref{Fried0B}), can be rewritten as
\begin{equation} \label{Fried1B}
\left(H^2+\frac{k}{a^2}\right)^{1-\Delta/2} = \frac{8\pi G_{\rm
eff}}{3}(\rho+\rho_{\Lambda}).
\end{equation}
Combining the modified Friedmann equation (\ref{Fried1B}) with the
continuity equation (\ref{Cont}), we arrive at
\begin{eqnarray}
&&(2-\Delta)\frac{\ddot{a}}{a}
\left(H^2+\frac{k}{a^2}\right)^{-\Delta/2}+(1+\Delta)\left(H^2+\frac{k}{a^2}\right)^{1-\Delta/2}
\nonumber
\\
&&=-8\pi G_{\rm eff}(p+p_{\Lambda}),\label{Fried2B}
\end{eqnarray}
where $p_{\Lambda}=-{\Lambda}/(8\pi G_{\rm eff})$. This is the
second modified Friedmann equation governing the evolution of the
universe based on Barrow entropy. In the limiting case where
$\Delta=0$ ($G_{\rm eff}\rightarrow G$), Eqs. (\ref{Fried1B}) and
(\ref{Fried2B}) reduce to the Friedmann equation in standard
cosmology.

In this case, we define the critical density parameter as
$\rho_{c}=3H^{2-\Delta}/(8\pi G_{\rm eff})$, and thus the first
Friedmann equation (\ref{Fried1B}) can be rewritten as
\begin{eqnarray}
\Omega_m+ \Omega_{\Lambda}=(1+\Omega_k)^{1-\Delta/2}.
\end{eqnarray}
Now we are going to show that the modified Friedmann equations
based on Tsallis/Barrow entropy derived in Eqs. (\ref{Fried1T})
and (\ref{Fried1B}) can describe the late time accelerated
expansion. For a flat universe filled with pressureless matter
($p=p_m=0$) and cosmological constant, we have $\Omega_m+
\Omega_{\Lambda}=1.$ The total equation of state parameter can be
written
\begin{eqnarray}
&&\omega_{\rm
t}=\frac{p_{\Lambda}}{\rho_m+\rho_{\Lambda}}=-\frac{1}{\rho_m/\rho_{\Lambda}+1}, \\
&&\Rightarrow\nonumber \omega_{\rm t}
(z)=-\frac{\Omega_{\Lambda,0}}{(1-\Omega_{\Lambda ,0})
(1+z)^3+\Omega_{\Lambda,0}},
\end{eqnarray}
where $\rho_{\Lambda}=\rho_{\Lambda,0}$, and $\rho_m=\rho_{m,0}
(1+z)^{3}$. If we take $\Omega_{\Lambda,0} \simeq 0.7 $ and
$\Omega_{m,0} \simeq 0.3$, we have
\begin{eqnarray}
&&\omega_{\rm t} (z)=-\frac{0.7}{0.7+0.3 (1+z)^3}.
\end{eqnarray}
At the present time where $z\rightarrow 0$, we have $\omega_{\rm
t}=-0.7$, while at the early universe where $z\rightarrow \infty$,
we get $\omega_{\rm t}=0$. This implies that at the early stages,
the universe undergoes a decelerated phase while at the late time
it experiences an accelerated phase.
\begin{figure}[t]
\epsfxsize=7.5cm \centerline{\epsffile{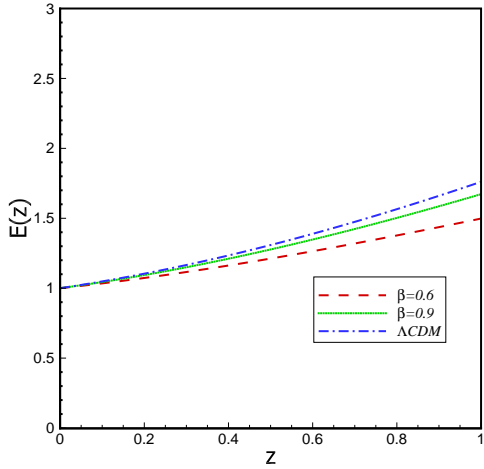}} \caption{The
behavior of $\omega_{\rm t}$ in terms of redshift $z$ for
different values of $\Omega_{\Lambda,0}$.} \label{Fig1}
\end{figure}
In Fig. 1 we have plotted the $\omega_{t}$ versus $z$ for
different values of $\Omega_{\Lambda,0}$. This figure confirms an
accelerated universe at the late time.
\section{Growth of perturbation in Tsallis cosmology\label{GTE}}
In a spatially flat universe, the modified  Friedmann equation
(\ref{Fried1T}) can be written as
\begin{eqnarray}
&& H^{4-2\beta}=\dfrac{1}{3}\left( \rho_{m}+\Lambda\right) ,
\label{Friedmann1}
\end{eqnarray}
where we have taken $8 \pi L_p^2 =1$ and hence
$\rho_{\Lambda}=\Lambda$. The second Friedmann equation
(\ref{Fried2T}) based on Tsallis entropy, is given by
\begin{eqnarray}
&& (4-2\beta)\dfrac{\ddot{a}}{a}H^{2-2\beta}+(2\beta-1)H^{4-2\beta}=\Lambda ,
\label{Friedmann2}
\end{eqnarray}
where $p_{m}=0$ and $p_{\Lambda}=-\Lambda$. The Hubble expansion
rate can be obtained as
\begin{equation}
H^{2}=\left( \dfrac{\rho_{m}+\Lambda}{3}\right)^{1/(2-\beta)}.
\label{Hubble}
\end{equation}
We can also define the normalized Hubble parameter as
\begin{equation}\label{Ez}
E(z)=\frac{H(z)}{H_{0}}=\left[
(1-\Omega_{\Lambda,0})(1+z)^{3}+\Omega_{\Lambda,0}\right]^{1/(4-2\beta)}.
\end{equation}
The evolution of the normalized Hubble parameter versus $z$ for
different values of $\beta$ is plotted in Fig. 2. As we can see,
in Tsallis cosmology the Hubble parameter is smaller than
$\Lambda$CDM model, implying that in this model, our Universe
expands slower. Also the Hubble parameter decreases with
decreasing the non-extensive parameter $\beta$.
\begin{figure}[t]
\epsfxsize=7.5cm \centerline{\epsffile{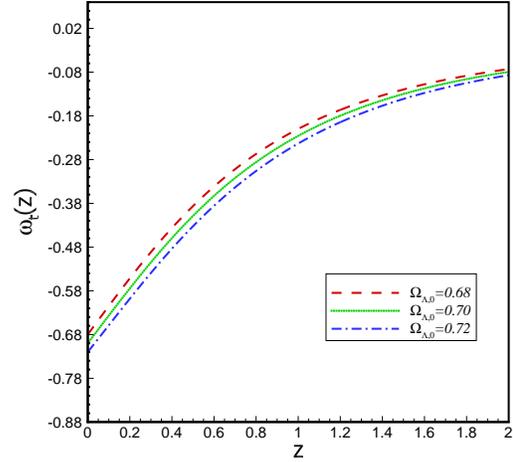}} \caption{The
behavior of the normalized Hubble rate $E(z)$ for different values
of $\beta$ in Tsallis cosmology, where we have taken $\Omega
_{\Lambda,0}=0.70$.} \label{Fig2}
\end{figure}
The deceleration parameter in terms of the redshift can be written
as
\begin{eqnarray}
&&q=-1-\dfrac{\dot{H}}{H^2}=-1+\dfrac{(1+z)}{H(z)}\dfrac{dH(z)}{dz} \nonumber\\
&&=-1+\dfrac{3}{2}\dfrac{(1+z)^{3}(1-\Omega_{\Lambda
,0})}{(2-\beta)[(1-\Omega_{\Lambda,0})(1+z)^{3}+\Omega_{\Lambda,0}]}.
\label{deceleration}
\end{eqnarray}
We have plotted the behavior of the deceleration parameter $q(z)$
for different $\beta$ parameter in Fig. 3. We observe that the
universe experiences a transition from a decelerating phase
($q>0$) to an accelerating phase ($q<0$), at redshift around
$z_{tr}\approx 0.6$ compatible with observations. It is seen that
$z_{tr}$ depends on the non-extensive parameter $\beta$ and
decreases with increasing $\beta$.
\begin{figure}[t]\label{Fig3}
\epsfxsize=7.5cm \centerline{\epsffile{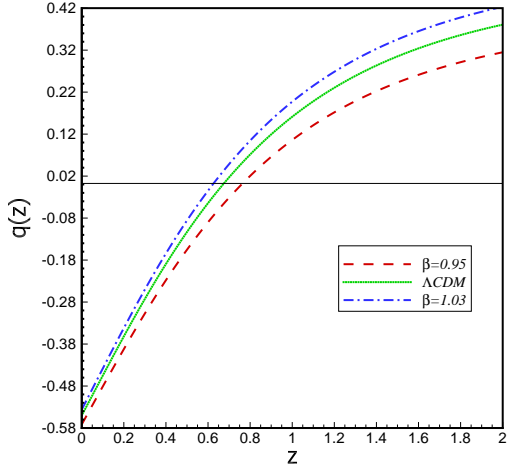}} \caption{The
behavior of the deceleration parameter $q(z)$ as a function of
redshift for different $\beta$ ($\beta=1$ corresponds to
$\Lambda$CDM model). Here we have taken $\Omega
_{\Lambda,0}=0.70$.}
\end{figure}

\begin{figure}[t]
\epsfxsize=7.5cm \centerline{\epsffile{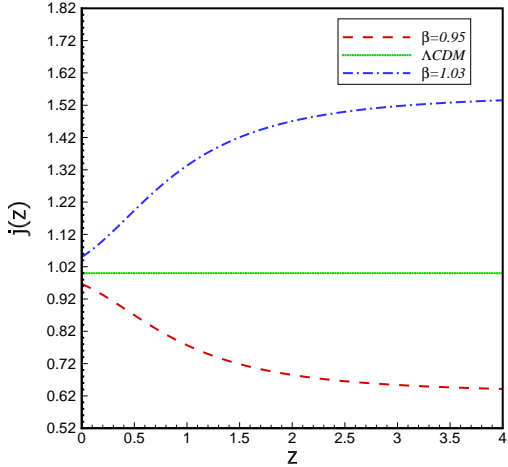}} \caption{The
evolution of jerk parameter with respect to redshift for different
values of $\beta$ parameter.} \label{Fig4}
\end{figure}

Another quantity which is helpful in understanding the phase
transitions of the universe is called the \textit{jerk parameter}.
This is a dimensionless quantity obtained by taking the third
derivative of the scale factor with respect to the cosmic time,
provides a comparison between different DE models and the
$\Lambda$CDM $(j=1)$ model. The jerk parameter is defined as
\begin{equation}
j=\dfrac{1}{aH^{3}}\dfrac{d^{3}a}{dt^{3}}=q(2q+1)+(1+z)\dfrac{dq}{dz}.
\label{jerk}
\end{equation}
For the $\Lambda$CDM model, the value of $j$ is always unity. A
non-$\Lambda$CDM model occurs if there is any deviation from the
value of $j=1$. From Fig. \ref{Fig4} we observe that the jerk
parameter is smaller (larger) than $\Lambda$CDM for $\beta<1$
($\beta>1$).
\

Combining Eqs. (\ref{Friedmann1}) and (\ref{Friedmann2}), we
arrive at
\begin{eqnarray}
&&\dfrac{\ddot{a}}{a}=\dfrac{1}{4-2\beta}\Bigg{\{}(1-2\beta)\left( \dfrac{\rho_{m}+\Lambda}{3}\right) ^{1/(2-\beta)} \nonumber\\
&&+\Lambda \left( \dfrac{\rho_{m}+\Lambda}{3}\right)
^{(\beta-1)/(2-\beta)}\Bigg{\}} . \label{adubbledot}
\end{eqnarray}
The conservation equations take the
form
\begin{equation}
\dot{\rho}_{m}+3H\rho_{m}=0.
\label{continutyDE1}
\end{equation}
This equation describes the density evolution of pressureless
matter labeled by $m$ with background density $\rho_{m}$. The SC
model describes a spherical region with uniform density so that at
time $t$, $\rho_{m}^{c}=\rho_{m} (t)+\delta\rho_{m}$. We can write
the conservation equation for spherical perturbed region of radius
$a_p$ as
\begin{equation}
\dot{\rho}_{m}^{c}+3h\rho_{m}^{c}=0,
\label{continutyDE2}
\end{equation}
where $h=\dot{a_{p}}/a_{p}$ is the local expansion rate of the
spherical perturbed region of radius $a_{p}$. Since Eq.
(\ref{adubbledot}) valid for all spacetime region, therefore we
further assume it holds for the spherical perturbed region with
radius $a_{p}$, namely,
\begin{eqnarray}\label{apdubbledot}
&&\dfrac{\ddot{a}_{p}}{a_{p}}=\dfrac{1}{4-2\beta}\Bigg{\{}
(1-2\beta)\left( \dfrac{\rho_{m}^{c}+\Lambda}{3}\right) ^{{1}/(2-\beta)} \nonumber\\
&&+\Lambda \left( \dfrac{\rho_{m}^{c}+\Lambda}{3}\right)
^{(\beta-1)/(2-\beta)}\Bigg{\}} ,
\end{eqnarray}
where for simplicity, we assume $\beta^{c}=\beta $. The density
contrast of a single fluid labeled by $\delta_m$, is defined as
\cite{Ziaei1}
\begin{equation}
\delta_{m}=\dfrac{\rho_{m}^{c}}{\rho_{m}}-1=\dfrac{\delta\rho_{m}}{\rho_{m}}.
\label{deltaDE}
\end{equation}
Taking the time derivative of Eq. (\ref{deltaDE}) and using Eqs.
(\ref{continutyDE1}) and (\ref{continutyDE2}), yields
\begin{equation}
\dot{\delta_{m}}=3(1+\delta_{m})(H-h). \label{deltadotDE}
\end{equation}
If we take the second derivative with respect to time, we find
\begin{equation}
\ddot{\delta}_{m}-\dfrac{\dot{\delta}_{m}^{2}}{1+\delta_{m}}=3(\dot{H}-\dot{h})(1+\delta_{m}), \label{deltadubbledotDE}
\end{equation}
where we have used Eq. (\ref{deltadotDE}). Combining Eqs.
(\ref{adubbledot}), (\ref{apdubbledot}) and (\ref{deltaDE}), we
can find the following equation
\begin{eqnarray}
\dot{H}-\dot{h}&=&-H^{2}+h^{2}+\dfrac{2\beta-1}{2(2-\beta)^{2}}
\left(\dfrac{\rho_{m}}{3}\right)^{1/(2-\beta)}\delta_{m} \nonumber\\
&&+\dfrac{\Lambda(1-\beta)}{2(2-\beta)^{2}} \left(
\dfrac{\rho_{m}}{3}\right)^{(\beta-1)/(2-\beta)}\delta_{m},\label{Hdot-hdot-DE}
\end{eqnarray}
where we have expanded the first and second terms in matter
dominated era and in the linear term of $\delta_{m}$. This is due
to the fact that we work in the early universe which
$\rho_{\Lambda}/\rho_{m}<1$ and in the linear regime where
$\delta_{m} <1$. Therefore, Eq. (\ref{deltadubbledotDE}) with
using Eqs. (\ref{deltadotDE}) and (\ref{Hdot-hdot-DE}) in linear
regime can be rewritten as
\begin{eqnarray}\label{deltadummledot2}
&&\ddot{\delta}_{m}+2H\dot{\delta}_{m}-\dfrac{3}{2}\dfrac{2\beta-1}{(2-\beta)^{2}}\left( \dfrac{\rho_{m}}{3}\right)^{1/(2-\beta)}\delta_{m}\nonumber\\
&&-\dfrac{3}{2}\dfrac{(1-\beta)\Lambda}{(2-\beta)^{2}}\left(
\dfrac{\rho_{m}}{3}\right)^{(\beta-1)/(2-\beta)}\delta_{m} =0,
\end{eqnarray}
where the matter energy density is given by $\rho_{m}
=\rho_{m,0}a^{−3}$. In order to study the evolution of the
density contrast $\delta_{m}$ in terms of the redshift parameter,
$1 + z = a^{-1}$, we first replace the time derivatives with the
derivatives with respect to the scale factor $a$. It is a matter
of calculations to show that
\begin{eqnarray}
&&\dot{\delta}_{m}=aH\delta^{\prime}_{m}, \nonumber\\
&&\ddot{\delta}_{m}=aH^{2}\left(\dfrac{1-2\beta+\Lambda
H^{2\beta-4}}{4-2\beta} \right) \delta^{\prime}_{m}
+a^{2}H^{2}\delta^{\prime\prime} _{m} \label{deltadotj2}
\end{eqnarray}
where the prime stands for the derivative with respect to the
scale factor $a$. Combining Eqs. (\ref{deltadummledot2}) and (\ref{deltadotj2}), we
arrive at
\begin{eqnarray}\label{Matterdeltaa3}
&&\delta^{\prime\prime}_{m}+\dfrac{1}{a(4-2\beta)}\left[(9-6\beta)+\Lambda H^{2\beta-4}\right] \delta^{\prime}_{m}\nonumber\\
&&+\Bigg{\{}(1-2\beta)\rho^{{1}/(2-\beta)}_{m}-3(1-\beta)\Lambda\rho^{(\beta-1)/(2-\beta)}_{m}\Bigg{\}}\nonumber\\
&&\times\dfrac{3^{(1-\beta)/(2-\beta)}}{a^2H^2}\dfrac{\delta_{m}}{2(2-\beta)^{2}}=0.
\label{DEdeltaa3}
\end{eqnarray}
It should be noted that in standard cosmology ($\beta=1$) and in
the absence of the cosmological constant ($\Lambda=0$), Eq.
(\ref{DEdeltaa3}) reduces to
\begin{equation}
\delta^{\prime\prime}_{m}+\dfrac{3}{2a}\delta^{\prime}_{m}-\dfrac{3}{2a^{2}}\delta_{m}=0,
\label{grt}
\end{equation}
which is the result obtained in standard cosmology \cite{Abramo}.
In Fig. 5, we plot the matter density contrast $\delta_m$ as a
function of redshift for different values of $\beta$ parameter and
for redshifts $10<z<100$.
\begin{figure}[t]
\epsfxsize=7.5cm \centerline{\epsffile{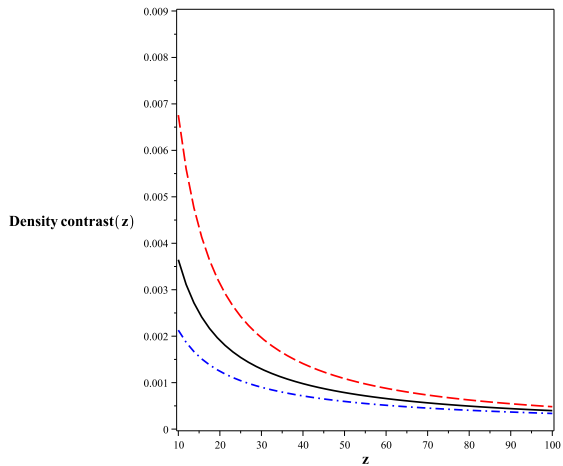}} \caption{Growth
of matter perturbations (density contrast $\delta_m$) for
different values of $\beta$ in Tsallis cosmology, where the
dashed-line for $\beta=1.06$, solid-line for $\Lambda$CDM
($\beta=1$), and dash-dotted line for $\beta=0.94$. Here we take
$\delta_{m}(z_{i})=0.0001$.} \label{Fig5}
\end{figure}
We can see that non-extensive parameter $\beta$ affects the growth
of matter perturbations, in particular in the lower redshifts.
Indeed the growth of perturbations increases with increasing the
non-extensive Tsallis parameter $\beta$. Further, we find out
that, in lower redshifts, the matter perturbations for $\beta<1$
grows slower compared to standard cosmology ($\beta=1$), while for
$\beta>1$ it grows up faster.
\section{Growth of perturbation in Barrow cosmology\label{GBE}}
In a flat FRW universe, the modified Friedmann equations, based on
Barrow entropy, can be rewritten
\begin{eqnarray}\label{FriedB1}
&&H^{2-\Delta}=\Gamma(\rho_{m}+\rho_{\Lambda}),\\
&&\left( 2-\Delta\right)\dfrac{\ddot{a}}{a}H^{-\Delta}+\left(
1+\Delta\right)
H^{2-\Delta}\nonumber\\&&=-3\Gamma(p_{m}+p_{\Lambda}),
\label{FriedB2}
\end{eqnarray}
where
\begin{equation}\Gamma
\equiv\dfrac{1}{3}\left(\dfrac{2-\Delta}{2+\Delta} \right)\left(
\dfrac{1}{8\pi^2}\right)^{{\Delta}/{2}}.
\end{equation}
Note that here we work in the unit where $k_{B}=c=\hslash=1$, and
thus $A_{0}=4G$. We also assume $8\pi G=1$, for simplicity. In
terms of the density parameters, the first Friedmann equation
(\ref{FriedB1}) can be written $\Omega_{m}+\Omega_{de}=1$, where
the dimensionless density parameters are now defined as
\begin{equation}\label{omegaBarr}
\Omega_{m}\equiv\dfrac{\Gamma\rho_{m}}{H^{2-\Delta}}, \   \   \  \
\ \Omega_{de}\equiv\dfrac{\Gamma\rho_{de}}{H^{2-\Delta}}.
\end{equation}
We can write the normalized Hubble parameter as
\begin{equation}
E(z)=\dfrac{H(z)}{H_{0}}=\left[
(1-\Omega_{\Lambda,0})(1+z)^{3}+\Omega_{\Lambda,0}\right]
^{1/(2-\Delta)}. \label{Ez2}
\end{equation}
The evolution of the normalized Hubble parameter versus $z$ for
different values of $\Delta$ is plotted in Fig. 6. It is seen that
the Hubble parameter is larger that $\Lambda$CDM $(\Delta=0)$
model, indicating that in Barrow cosmology the universe expands
faster. Also, the Hubble parameter decreases with decreasing the
Barrow parameter $\Delta$.
\begin{figure}[t]
\epsfxsize=7.5cm \centerline{\epsffile{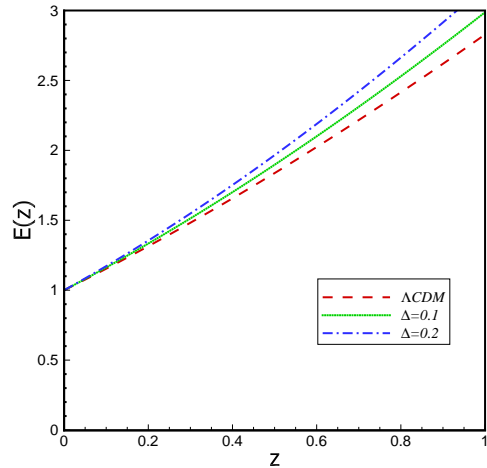}} \caption{The
behavior of the normalized Hubble rate for different values of
$\Delta$ where we have taken $\Omega_{\Lambda,0}=0.7$.}
\label{Fig6}
\end{figure}\\
The deceleration parameter in terms of the redshift can be written as
\begin{eqnarray}
&&q=-1-\dfrac{\dot{H}}{H^2}=-1+\dfrac{(1+z)}{H(z)}\dfrac{dH(z)}{dz} \nonumber\\
&&=-1+\dfrac{3(1+z)^{3}(1-\Omega_{\Lambda,0})}{(2-\Delta)\left((1-\Omega_{\Lambda,0})(1+z)^{3}+\Omega_{\Lambda,0}
\right) }. \label{deceleration2}
\end{eqnarray}
We have plotted the behavior of the deceleration parameter $q(z)$
for different $\Delta$ parameter in Fig. 7. We observe that the
universe experiences a transition from decelerating phase
($z>z_{tr}$) to accelerating phase ($z<z_{tr}$). Also we can see
that with increasing $\Delta$, the phase transition between
deceleration and acceleration take place at
lower redshifts.\\
\begin{figure}[t]
\epsfxsize=7.5cm \centerline{\epsffile{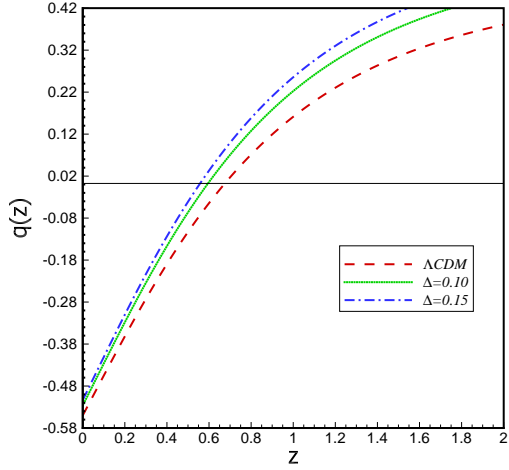}} \caption{The
behavior of the deceleration parameter as a function of redshift
for different $\Delta$. Here we take $\Omega _{\Lambda,0}=0.7$.}
\label{Fig7}
\end{figure}
Similar to previous section, we can find the \textit{jerk
parameter} in this model. From Fig. 8 we observe the jerk
parameter for all values of $\Delta$ parameter is greater than
$\Lambda$CDM model.
\begin{figure}[t]
\epsfxsize=7.5cm \centerline{\epsffile{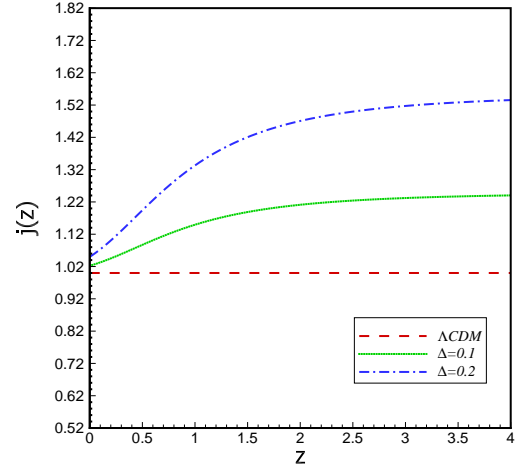}} \caption{The
evolution of jerk parameter with respect to redshift for different
values of $\Delta$ parameter.} \label{Fig8}
\end{figure}\
Combining Eqs. (\ref{FriedB1}) and (\ref{FriedB2}), we find
\begin{eqnarray}
&&\dfrac{\ddot{a}}{a}=\frac{\Gamma^{2/(2-\Delta)}}{\Delta-2}\Bigg{\{}
  (1+\Delta)\left( \rho_{m}+\Lambda\right)^{2/(2-\Delta)} \nonumber\\
&&-3\Lambda\left( \rho_{m}+\Lambda\right)
^{\Delta/(2-\Delta)}\Bigg{\}}. \label{adubbledotBarr}
\end{eqnarray}
In order to study the growth of
perturbations in Barrow cosmology, following the approach of the
previous section, we employ the SC method. We can write down for
the spherical perturbed region of radius $a_p$,
\begin{eqnarray}\label{apdubbledotBarr}
&&\frac{\ddot{a}_{p}}{a_{p}}=\frac{\Gamma^{2/(2-\Delta)}}{\Delta-2}\Bigg{\{}
 (1+\Delta)\left( \rho_{m}^{c}+\Lambda\right) ^{2/(2-\Delta)} \nonumber\\
&&-3\Lambda\left(\rho_{m}^{c}+\Lambda\right)
^{\Delta/(2-\Delta)}\Bigg{\}} ,
\end{eqnarray}
Combining Eqs. (\ref{deltaDE}), (\ref{adubbledotBarr}) and
(\ref{apdubbledotBarr}), we arrive at
\begin{eqnarray}
&&\dot{H}-\dot{h}+H^{2}-h^{2}=\Bigg{\{}\dfrac{2(1+\Delta)}{(2-\Delta)^{2}}(\Gamma\rho_{m})^{2/(2-\Delta)}\nonumber\\
&&-\dfrac{3\Delta}{(2-\Delta)^{2}}\Lambda(\Gamma\rho_{m})^{\Delta/(2-\Delta)}\Bigg{\}}\delta_{m}.
\label{Hdot-hdot-DE-Barr}
\end{eqnarray}
Using Eqs. (\ref{deltadotDE}), (\ref{deltadubbledotDE}) and
(\ref{Hdot-hdot-DE-Barr}), after a simple calculations, we find
\begin{eqnarray}\label{Matterdelta1Barr}
&&\ddot{\delta}_{m}+2H\dot{\delta}_{m}-\Bigg{\{}\dfrac{6(1+\Delta)}{(2-\Delta)^{2}}(\Gamma\rho_{m})^{2/(2-\Delta)}\nonumber\\
&&-\dfrac{9\Delta}{(2-\Delta)^{2}}\Lambda(\Gamma\rho_{m})^{\Delta/(2-\Delta)}\Bigg{\}}\delta_{m}=0.
 \label{DEdelta1Barr}
\end{eqnarray}
Changing our variable from time to the scale factor, we can
rewrite the above equations as
\begin{eqnarray}\label{Matterdeltaa3Barr}
&&\delta^{\prime\prime}_{m}+\dfrac{3}{a(2-\Delta)}\left[(1-\Delta)+\Gamma\Lambda H^{\Delta-2}\right] \delta^{\prime}_{m}\nonumber\\
&&-\Bigg{\{}2(1+\Delta)(\Gamma\rho_{m})^{2/(2-\Delta)}-3\Delta\Lambda(\Gamma\rho_{m})^{\Delta/(2-\Delta)}\Bigg{\}}\nonumber\\
&&\times\dfrac{3}{a^2H^2}\dfrac{\delta_{m}}{(2-\Delta)^{2}}=0.
\label{DEdeltaa3Barr}
\end{eqnarray}
It should be noted that in standard cosmology ($\Delta=0$) and
with the absence of the cosmological constant ($\Lambda = 0$),
this equation reduces to Eq. (\ref{grt}). In Fig. 9, we have
plotted the matter density contrast as a function of redshift for
different values of $\Delta$ parameter and for redshifts $10 < z <
100$.
\begin{figure}[t]
\epsfxsize=7.5cm \centerline{\epsffile{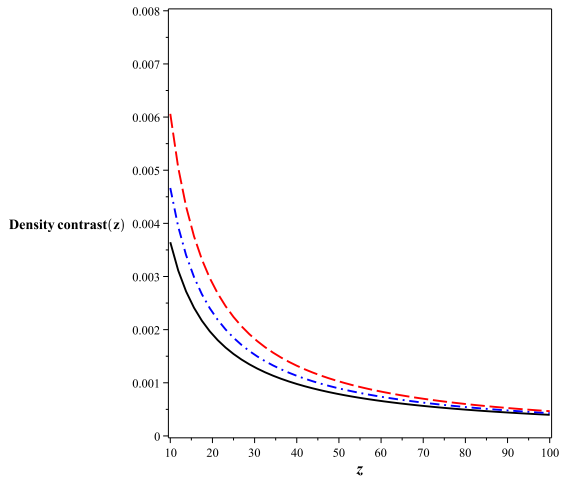}} \caption{Growth
of matter perturbations (density contrast $\delta_m$) for
different values of $\Delta$, where dashed-line for $\Delta=0.1$,
dash-dotted line for $\Delta=0.05$ and solid-line for $\Lambda
CDM$. Here we take $\delta_{m}(z_{i})=0.0001$} \label{Fig9}
\end{figure}
We can see that Barrow  exponent $\Delta$ affects the growth of
total perturbations, in particular in the lower redshifts. Indeed,
in low redshifts, the total density contrast increases with
increasing $\Delta$ parameter. This implies that in Barrow
cosmology, the perturbations grow up faster compared to standard
cosmology. This is due to the fact that in Barrow cosmology, the
spacetime has quantum-gravitational deformation, and intricate
fractal structure, which can support the growth of perturbations
in energy density.
\section{Concluding remarks\label{Con}}
In this paper, we have investigated the modified cosmological
field equations of flat FRW universe when the entropy associated
with the apparent horizon is in the form of Tsallis/Barrow
entropy. We have explored the growth of perturbations in the
context of Tsallis/Barrow cosmology. We have considered the
energy/matter content of the universe in the form of matter
(baronic matter and DM) and cosmological constant. Employing the
SC approach for the growth of perturbations, we extracted the
differential equation for the evolution of the matter
perturbations. Further, for the matter density contrast, we
observed that the profile of the growth of perturbations differ
from standard cosmology. Indeed, in Tsallis cosmology and in lower
redshifts, the perturbations grow slower for $\beta<1$ compared to
the standard cosmology ($\beta=1$), while they grow faster for
$\beta>1$. In Barrow cosmology, however, the growth rate of the
total perturbations always became faster compared to the standard
cosmology.

Also we find that both Tsallis and Barrow cosmology can explain
the phase transition from a decelerating to an accelerating
universe. In contrast to Tsallis cosmology, we observed that
Barrow cosmology predict that the universe expands with greater
rate compared to $\Lambda$CDM model.

It is important to note that although the modified Friedmann
equations based on Tsallis and Barrow entropy are similar to each
other, but the growth of perturbation differs in some cases in
these models. This is due to the fact that the non-extensive
Tsallis parameter $\beta$ takes different value from Barrow
exponent $\Delta$. Indeed  non-extensive Tsallis parameter has an
upper bound, $\beta<2$, while Barrow exponent ranges as
$0\leq\Delta\leq1$. Besides, the origin of these two corrections
to the entropy are completely different.

Finally, it is worth mentioning that Barrow entropy correction to
the area law is just a first approximation on the subject of
quantum gravitational implications of the black hole horizons. In
reality the underlying spacetime foam deformation will be complex,
wild and dynamical. So one could think of an exponent $\Delta$
that depends on time and scale, as it has already been done with
Tsallis entropy exponent \cite{S. Nojiri}. In order to constrain
the Barrow exponent $\Delta$ and Tsallis nonextensive parameter
$\beta$, one may perform a full observational confrontation using
data from Supernovae type Ia data (SNIa), Cosmic Microwave
Background (CMB) shift parameters, Baryonic Acoustic Oscillations
(BAO), growth rate and Hubble data observations. These issues are
beyond the scope of the present work and we leave them for the
future investigations.
\acknowledgments{We are grateful to the referee for very helpful
and constructive comments which helped us improve our paper
significantly.}


\begin{thebibliography}{99}

\bibitem{Peebles} P. Peebles, \textit{Principles of Physical Cosmology}, Princeton University Press, (Princeton, NJ, 1993).

\bibitem{White} S. D. M. White and M. J. Rees, \textit{Core condensation in heavy halos:
a two-stage theory for galaxy formation and clustering}, Mon. Not.
R. Astron. Soc {\bf183}, 341 (1978).

\bibitem{Abramo} R. Abramo, R. Batista, L. Liberato and R.Rosenfeld, \textit{Struture formation in the presene of dark
energy perturbations}, JCAP {\bf 11}, 012 (2007)
[arXiv:0707.2882].

\bibitem{Abramo2} L. Abramo, R. Batista, L. Liberato, and R. Rosenfeld,
\textit{Physical approximations for the nonlinear evolution of perturbations in dark energy scenarios}, Phys. Rev. D {\bf79}, 023516 (2009) [arXiv:0806.3461].

\bibitem{Nunes} N. J. Nunes, A. C. da Silva and N. Aghanim, \textit{Number counts in homogeneous and inhomogeneous dark energy models}, Astron. Astrophys {\bf450}, 899 (2005) [arXiv:astro-ph/0506043].

\bibitem{Liberato} L. Liberato and R. Rosenfeld, \textit{Dark energy parametrizations and their effect on dark halos}, JCAP {\bf07}, 009 (2006) [arXiv:astro-ph/0604071].


\bibitem{Dutta} S. Dutta and I. Maor, \textit{Voids of dark energy}, Phys. Rev. D {\bf75}, 063507 (2007) [arXiv:gr-qc/0612027].

\bibitem{Amendola} L. Amendola and S. Tsujikawa, \textit{Dark energy-Theory and Observations}, Cambridge University Press (2010).

\bibitem{Planelles} S. Planelles, D. Schleicher, A. Bykov,
\textit{Large- scale structure formation: from the first
non-linear objects to massive galaxy clusters}, Space Sci. Rev
{\bf51}, 93 (2016) [arXiv:1404.3956].

\bibitem{Ziaei1} A. Ziaie, H. Shabani, S. Ghaffari, \textit{Effects of Rastall parameter on perturbation of dark sectors of the Universe}, Mod. Phys. Lett. A {\bf36}, 2150082 (2021) [arXiv:1909.12085].
\bibitem{Ziaei2} A. H. Ziaie, H. Moradpour, H. Shabani, \textit{Structure Formation in Generalized Rastall Gravity}, Eur. Phys. J. Plus {\bf 135}, 916 (2020) [arXiv:2002.12757].
\bibitem{Farsi} B. Farsi and A. Sheykhi, \textit{Structure formation in mimetic gravity}, [arXiv:2202.04118v1].

\bibitem{Jacobson} T. Jacobson, \textit{Thermodynamics of space-time: The Einstein equation of state}, Phys. Rev. Lett {\bf75}, 1260 (1995) [arXiv:gr-qc/9504004].
\bibitem{Padmanabhan1} T. Padmanabhan, \textit{Classical and quantum thermodynamics of horizons
in spherically symmetric spacetimes}, Classical and Quantum
Gravity {\bf19}, 5387 (2002) [arXiv:gr-qc/0204019].
\bibitem{Padmanabhan2} T. Padmanabhan, \textit{Gravity and the thermodynamics of horizons}, Phys. Rept {\bf406}, 49 (2005)[arXiv:gr-qc/0311036].
\bibitem{Paranjape} A. Paranjape, S. Sarkar, and T. Padmanabhan, \textit{Thermodynamic route to field equations in Lanczos-Lovelock gravity}, Phys. Rev. D {\bf74}, 104015 (2006) [arXiv:hep-th/0607240].
\bibitem{Kothawala} D. Kothawala, S. Sarkar, and T. Padmanabhan, \textit{Einstein field equations as a thermodynamic identity: The cases of stationary axisymmetric horizons and evolving spherically symmetric horizons}, Phys. Lett. B {\bf652(5) }, 338 (2007) [arXiv:gr-qc/0701002].

\bibitem{Eling} C. Eling, R. Guedens, and T. Jacobson, \textit{Non-equilibrium thermodynamics of spacetime}, Phys. Rev. Lett  {\bf96}, 121301 (2006) [arXiv:gr-qc/0602001].
\bibitem{Padmanabhan3} T. Padmanabhan, \textit{Thermodynamical Aspects of Gravity: New insights}, Rept. Prog. Phys
{\bf73}, 046901 (2010) [arXiv:0911.5004].
\bibitem{Frolov} A. V. Frolov and L. Kofman, \textit{Inflation and de Sitter thermodynamics}, JCAP {\bf0305}, 009 (2003) [arXiv:hep-th/0212327].
\bibitem{Calcagni} G. Calcagni, \textit{de Sitter thermodynamics and the braneworld}, JHEP {\bf0509}, 060 (2005) [arXiv:hep-th/0507125].
\bibitem{Cai1} R.G. Cai and S.P. Kim, \textit{First law of thermodynamics and Friedmann equations of Friedmann-Robertson-Walker universe}, JHEP {\bf02}, 050 (2005) [arXiv:hep-th/0501055].
\bibitem{Cai2}R. G. Cai and N. Ohta, \textit{Horizon Thermodynamics and Gravitational Field Equations in Horava-Lifshitz Gravity}, Phys. Rev. D {\bf81}, 084061 (2010) [arXiv:0910.2307]
\bibitem{Akbar1}M. Akbar and Rong-Gen Cai, \textit{Friedmann equations of FRW universe in scalar-tensor gravity, f(R) gravity and first law of thermodynamics}, Phys. Lett. B {\bf635(1) }, 7 (2006) [arXiv:hep-th/0602156].
\bibitem{Akbar2} M. Akbar and R. G. Cai, \textit{Thermodynamic Behavior of Friedmann Equations at Apparent Horizon of FRW Universe}, Phys. Rev. D {\bf75}, 084003 (2007) [arXiv:hep-th/0609128].
\bibitem{Akbar3} M. Akbar and R. G. Cai, \textit{Friedmann equations of FRW universe in scalar-tensor gravity, f(R) gravity and first law of thermodynamics}, Phys. Lett. B {\bf635}, 7 (2006) [arXiv:hep-th/0602156].
\bibitem{Sheykhi1} A. Sheykhi, \textit{Modified Friedmann equations from Tsallis entropy}, Phys. Lett. B {\bf785}, 118 (2018) [arXiv:1806.03996] .
\bibitem{Sheykhi2} A. Sheykhi, \textit{Barrow Entropy Corrections to Friedmann Equations}, Phys. Rev. D {\bf103}, 123503 (2021) [arXiv:2102.06550].
\bibitem{Sheykhi3} A. Sheykhi, B. Wang and R. G. Cai, \textit{Thermodynamical Properties of Apparent Horizon in Warped DGP Braneworld}, Nucl. Phys. B {\bf779}, 1 (2007) [arXiv:hep-th/0701198].
\bibitem{Sheykhi4} A. Sheykhi, B. Wang, and R. Cai, \textit{Deep connection between thermodynamics and gravity in Gauss-Bonnet braneworlds}, Phys. Rev. D {\bf76}, 023515 (2007) [arXiv:hep-th/0701261].
\bibitem{Sheykhi5} A. Sheykhi and B. Wang, \textit{Generalized second law of thermodynamics in Gauss-Bonnet braneworld}, Physics Letters B {\bf678(5) }, 434 (2009) [arXiv:0811.4478].
\bibitem{SheyECFE} A. Sheykhi, \textit{Entropic corrections to Friedmann equations}, Phys. Rev. D {\bf81}, 104011 (2010), [arXiv:1004.0627].
\bibitem{Sheyem} A. Sheykhi, \textit{Friedmann equations from emergence of cosmic
space}, Phys. Rev. D {\bf87}, 061501(R) (2013), [arXiv:1304.3054].

\bibitem{Das} S. Das, P. Majumdar, R. K. Bhaduri, \textit{General logarithmic corrections to black-hole entropy}, Class. Quant  Grav {\bf19}, 2355 (2002) [arXiv:hep-th/0111001].

\bibitem{Ashtekar} A. Ashtekar, J. Baez, A. Corichi, and K. Krasnov, \textit{Quantum Geometry and Black Hole Entropy}, Phys. Rev. Lett {\bf80}, 904 (1998) [arXiv:gr-qc/9710007].
\bibitem{Zhang} J. Zhang, \textit{Black hole quantum tunnelling and black hole entropy correction}, Phys. Lett. B {\bf668}, 353 (2008) [arXiv:0806.2441].

\bibitem{Banerjee} R. Banerjee and B. R. Majhi, \textit{Quantum Tunneling and Back Reaction}, Phys. Lett. B {\bf662}, 62 (2008) [arXiv:0801.0200].
\bibitem{SheyLog} A. Sheykhi, \textit{Thermodynamics of apparent horizon and modified Friedmann equations},
Eur. Phys. J. C {\bf69}, 265 (2010), [arXiv:1012.0383].
\bibitem{Das2} S.Das, S. Shankaranarayanan, and S. Sur, \textit{Power-law corrections to entanglement entropy of horizons}, Phys. Rev. D {\bf77}, 064013 (2008) [arXiv:0705.2070].
\bibitem{Radicella} N. Radicella and D. Pavon, \textit{The generalized second law in universes with quantum corrected entropy relations}, Phys. Lett. B {\bf691(3) }, 121 (2010) [arXiv:1006.3745].
\bibitem{SheyPL} A. Sheykhi and S. H. Hendi, \textit{Power-law entropic corrections to Newton law and Friedmann equations},
Phys. Rev. D {\bf84}, 044023 (2011), [arXiv:1011.0676].

\bibitem{Wilk} G. Wilk and Z. Wlodarczyk, \textit{Interpretation of the Nonextensivity Parameter $q$ in Some Applications of Tsallis Statistics and Levy Distributions}, Phys. Rev. Lett {\bf84}, 2770 (2000) [arXiv:hep-ph/9908459].
\bibitem{Gibbs} J. Gibbs, \textit{Elementary Principles in Statistical Mechanics: Developed with Especial Reference to the Rational Foundation of Thermodynamics}, Cambridge Library Collection - Mathematics, (Cambridge University Press, 2010).

\bibitem{RNunes} R. Nunes, M. Barboza Jr., E. Abreu, and J. Ananias Neto, \textit{Dark energy models through nonextensive Tsallis statistics}, (2014) [arXiv:1403.5706].
\bibitem{Tsallis1} C. Tsallis, \textit{Possible generalization of Boltzmann-Gibbs statistics}, J. Stat. Phys {\bf52}, 479 (1988).
\bibitem{Tsallis2} C. Tsallis, L. J. L. Cirto, \textit{Black hole thermodynamical entropy}, Eur. Phys. J. C {\bf73}, 2487 (2013) [arXiv:1202.2154].


\bibitem{Czinner}V. G. Czinner, H. Iguchi, \textit{Thermodynamics, stability and Hawking Page transition of Kerr black holes from Rnyi statistics}, Eur. Phys. J. C {\bf77}, 892 (2017) .

\bibitem{Sayahian} A. Sayahian Jahromi et al., \textit{Generalized entropy formalism and a new holographic dark energy model}, Phys. Lett. B {\bf780}, 21 (2018) .

\bibitem{Moradpour} H. Moradpour et al.,\textit{Thermodynamic approach to holographic dark energy and the Renyi entropy}, Eur. Phys. J. C {\bf78}, 829 (2018) [arXiv:1803.02195].

\bibitem{Asghari} M. Asghari, A. Sheykhi, \textit{Observational constraints on Tsallis modified gravity}, MNRAS {\bf508}, 2855 (2021) [arXiv:2106.15551].
\bibitem{Tavayef} M. Tavayef, A. Sheykhi, K. Bamba, H. Moradpour, \textit{Tsallis holographic dark energy}, Physics Letters B {\bf781}, 195 (2018) [arXiv:1804.02983].
\bibitem{Abd} M. Abdollahi Zadeh, A. Sheykhi, H. Moradpour, K.
Bamba, \textit{A Note on Tsallis Holographic Dark Energy}, Eur.
Phys. J. C {\bf78}, 940 (2018) [arXiv:1806.07285].

\bibitem{Bram} B. D.Pandey, et. al., \textit{New Tsallis Holographic Dark
Energy}, Eur. Phys. J. C {\bf82}, 233 (2022), [arXiv:2110.13628].
\bibitem{Huang} Qihong Huang, He Huang, Jun Chen, Lu Zhang and Feiquan Tu, \textit{Stability analysis of the Tsallis holographic dark energy model}, Class. Quantum Grav {\bf36}, 175001(2019), [arXiv:2201.12504].

\bibitem{Bhattacharjee} S. Bhattacharjee, \textit{Growth rate and configurational entropy in Tsallis holographic dark energy}, Eur. Phys. J. C {\bf81}, 217 (2021) [arXiv:2011.13135].

\bibitem{Barrow} J. D. Barrow, \textit{The area of a rough black hole}, Phys. Lett. B {\bf808}, 135643 (2020) [arXiv:2004.09444].
\bibitem{Saridakis1} E.N. Saridakis, \textit{Modified cosmology through spacetime thermodynamics and Barrow horizon entropy}, JCAP {\bf07}, 031 (2020) [arXiv:2006.01105].
\bibitem{Saridakis2}E. N. Saridakis and S. Basilakos, \textit{The generalized second law of thermodynamics with Barrow entropy}, Eur. Phys. J. C {\bf81}, 644 (2021) [arXiv:2005.08258].
\bibitem{Saridakis3} E. N. Saridakis, \textit{Barrow holographic dark energy}, Phys. Rev. D {\bf 102}, 123525 (2020) [arXiv:2005.04115].
\bibitem{Sri} S. Srivastava, U. Kumar Sharma, \textit{Barrow holographic dark energy with Hubble horizon as IR cutoff}, Int. J. Geom. Methods Mod. Phys {\bf18(1) }, 2150014
(2021) [arXiv:2010.09439].

\bibitem{Adh} P. Adhikary, S. Das, S. Basilakos, E. N. Saridakis, \textit{Barrow Holographic Dark Energy in non-flat
Universe}, Phys. Rev. D {\bf104}, 123519 [arXiv:2104.13118].
\bibitem{Oliv} A. Oliveros, M. A. Sabogal, Mario A. Acero, \textit{Barrow holographic dark energy with Granda-Oliveros
cut-off}, [arXiv:2203.14464].

\bibitem{Anagnostopoulos} F. K. Anagnostopoulos, S. Basilakos and E. N. Saridakis,
\textit{Observational constraints on Barrow holographic dark energy}, Eur. Phys. J. C {\bf80}, 826 (2020) [arXiv:2005.10302].
\bibitem{Dabrowski} M. P. Dabrowski and V. Salzano,
\textit{Geometrical observational bounds on a fractal horizon holographic dark energy}, Phys. Rev. D {\bf102}, 064047 (2020) [arXiv:2009.08306].

\bibitem{Mamon} A. A. Mamon, A. Paliathanasis and S. Saha,
\textit{Dynamics of an Interacting Barrow Holographic Dark Energy Model and its Thermodynamic Implications}, Eur. Phys. J. Plus {\bf136}, 134 (2021) [arXiv:2007.16020].
\bibitem{Hay} S. A. Hayward, S.Mukohyana, and M. C. Ashworth, Phys. Lett. A {\bf 256}, 347 (1999).
\bibitem{S. Nojiri} S. Nojiri, S. D. Odintsov, E. N. Saridakis and R. Myrzakulov,
\textit{Correspondence of cosmology from non-extensive
thermodynamics with fluids of generalized equation of state},
Nucl. Phys. B {\bf950}, 114850 (2020) [arXiv:1911.03606].
\end{thebibliography}
\end{document}